\documentclass[10pt]{article}

\usepackage{graphics}
\usepackage{graphicx}

\usepackage[a4paper, left=35mm,right=35mm,top=34mm,bottom=34mm]{geometry}
\usepackage[utf8]{inputenc}
\usepackage[T1]{fontenc}
\usepackage[english]{babel}

\usepackage{enumerate}
\usepackage{graphicx}
\usepackage{hyperref}
\hypersetup{
    colorlinks=true,
    linkcolor=blue,
    filecolor=magenta,      
    urlcolor=cyan,
}
\usepackage{listings}
\usepackage{color}

\definecolor{dkgreen}{rgb}{0,0.6,0}
\definecolor{gray}{rgb}{0.5,0.5,0.5}
\definecolor{mauve}{rgb}{0.58,0,0.82}
\lstdefinelanguage{MRGC++}{%
  language=C++,
  morekeywords={T, U, MPI_Irecv, MPI_Isend, MPI_Allreduce, MPI_Waitall, Compute, Map, abs, max, Swap, MPI_Recv_init, MPI_Send_init, MPI_Startall, Copy, Init, InitRecv, InitSend, InitAllReduce, Send, Recv, AllReduce, Finalize, InitSnapshot, Snapshot, SwitchAsync, SnapReduce, MPI_Test, MPI_Start}
}
\usepackage{mathtools,amsthm,amssymb,amsfonts}
\usepackage{algorithm}
\usepackage{algorithmic}
\makeatother
\theoremstyle{plain}

\theoremstyle{definition}

\theoremstyle{remark}

\usepackage{caption} 
\usepackage{fancyhdr}

\author{
  {\normalsize Guillaume Gbikpi-Benissan}\thanks{CentraleSup\'elec, Universit\'e Paris Saclay, France
    (correspondence, frederic.magoules@hotmail.com).}
  \and
  {\normalsize Patrick Callet}\footnotemark[1]
  \and
  {\normalsize Fr\'ed\'eric Magoul\`es}\footnotemark[1]
}
\title{Spectral domain decomposition method for physically-based rendering of Royaumont abbey}
\date{}

\begin{document}
\maketitle
\thispagestyle{fancy}

\begin{abstract}
\noindent In the context of a virtual reconstitution of the destroyed Royaumont abbey church, this paper investigates computer sciences issues intrinsic to the physically-based image rendering. First, a virtual model was designed from historical sources and archaeological descriptions. Then some materials physical properties were measured on remains of the church and on pieces from similar ancient churches. We specify the properties of our lighting source which is a representation of the sun, and present the rendering algorithm implemented in our software Virtuelium. In order to accelerate the computation of the interactions between light-rays and objects, this ray-tracing algorithm is parallelized by means of domain decomposition techniques. Numerical experiments show that the computational time saved by a classic parallelization is much less significant than that gained with our approach.
\end{abstract}


\section{Introduction} \label{sec:introduction}

This study aims to propose a new computer science method for increasing the efficiency of physically-based image rendering, in terms of accuracy and computational time. It is a part of the rehabilitation project led on the Royaumont abbey which church edifice was destroyed in 1792. The physically-based image rendering is used to simulate a realistic visual aspect of the interior of the church according to natural lighting and medieval glasses properties.
Ancient churches form a part of the Christian cultural heritage in Europe. It is also of scientific interest to study and understand ancient techniques in art, particularly architectural properties of such edifices, like structural mechanics, acoustics and optics. Being able to elaborate and work on virtual models is a valuable asset for physically-based simulation. Unfortunately, some of these churches have been destroyed and therefore usual methods like points cloud acquisition~\cite{KBKHL2004,add1,add2} can not be used to build their associated Computer Aided Design (CAD) models. It is the case of the Royaumont abbey which is of interest in this study and therefore collaborative work between archaeologists, historians and computer scientists is mandatory to build a realistic and accurate model.

In this paper our goal is to simulate high quality visual appearance within the church while taking into account interactions between natural lighting and objects materials.
Building a virtual scene in image rendering requires the description of three main actors:
(i) the objects shapes and materials to be rendered,
(ii) the scene lighting, in order to make the objects visible and/or to enhance them,
(iii) the observer (usually the camera's position and properties) from which the scene is rendered.

With a physically-based rendering engine, the scene description is built from spectral data. Although this kind of description is more complex than the trichromatic one, traditionally used in computer graphics, such an approach brings many advantages. Indeed, the way photochromic glasses can change their shade depending on environment factors can only be understood if their physical micro-structures are known. Besides, their behavior cannot be separated from the nature of the light they filter. Thus, the visual adaptation of the observer to the lighting simulated environment requires colorimetric considerations. All these reasons make physically-based rendering a suitable approach if an accurate simulation is to be conducted. Well known rendering methods~\cite{Pharr:2010:PBR:1854996, Shirley:2012:BPR:2407783.2407785} are based on extrinsic properties of materials, i.e., reflectance or transmittance properties only. The research presented in this paper has been done using a multi-spectral physically-based ray-tracing software developed to render the visual appearance of materials and able to do multi-scaled simulations of light interactions.

In the following we first give the context in which the church edifice was modeled, including some general presentation of the Royaumont abbey history. Secondly, we described some requirements of the physically-based rendering process and particular aspects we considered for our simulation. Thirdly the rendering algorithm, mainly based on ray-tracing, is presented after the equations describing interactions between light-rays and objects inside the three-dimensional scene. Fourthly, our original parallelization method, based on domain decomposition, for accelerating the rendering process is described. Finally, we illustrate some results obtained according to some renderings of the interior of the church.

\section{Virtual reconstruction of the church of Royaumont abbey}

The Royaumont abbey is a royal edifice built from 1228 to 1235 at approximately 35 km in the north of Paris (France). King Saint Louis\footnote{1214-1270} and his mother, Blanche de Castille\footnote{1188-1252}, made of this Cistercian monastery a very important one in Europe. But over the centuries and wars, from about 140 monks at the death of Saint Louis, the abbey housed less than 30 monks at the end of the 18th century. After multiple alterations, as reported by historians, a main destruction occurred, after the 1789 French revolution, in 1792 when a cotton manufacturer demolished the church and utilized its stones to put up lodgings for workers. During the 20th century, damaged portions of the abbey were restored but very few parts of the church (some columns, a wall and a tower) remain today. Established in 1964, the Fondation Royaumont tries to maintain the historical, religious and architectural heritage from the remaining parts of the abbey, which was classified as a historic monument in France in 1927.

With so few physical elements, making up a virtual model of the church by computer scientists required collaborative work with archaeologists, who deduced a more complete aspect of the edifice, and with historians, who analyzed historical descriptions. One of the most accomplished historical source is due to the work of Millin~\cite{Mil1791} in 1791. Some engravings and complementary archaeological sources were also used in addition to the architectural schemes of Louis Vernier. Then, many inconsistencies, but also missing information, were analyzed and fixed. Fig.~\ref{fig:globmodel} illustrates the assembly of each little piece of the edifice we modeled.

\section{Physically-based rendering}

In order to simulate a high quality of the church, a certain amount of physical data should be provided to an appropriate rendering software, here Virtuelium. Properties of all objects materials are important since they influence the internal scattering of natural light inside the edifice. For instance, with the same exterior lighting, the interior of the church may look darker or over-illuminated, depending on the light absorption factor of its walls. Especially for window glasses, we determined the complex index of refraction using spectroscopic ellipsometry. This index is defined by (see~\cite{palik1985handbook})
\begin{equation}
\label{eq:optical}
\tilde{n}(\lambda) = n(\lambda)+ i k(\lambda) = n(\lambda) (1 + i \kappa(\lambda))
\end{equation}
where $i$ is the imaginary complex number, $n(\lambda)$ denotes the optical index, $\kappa(\lambda)$ denotes the index of absorption and $\lambda$ denotes the wavelength of light in vacuum. 
As it can be seen in~\cite{callet1998couleur}, these optical constants are required to define a highly realistic optical behavior of materials. Despite spectroscopic ellipsometry is well suited to define these optical constants for materials satisfying the Fresnel conditions (non-scattering and homogeneous) like modern glasses, it is not the case for medieval glasses because of the irregularity of the surface and the heterogeneity of the glasses. As a consequence, spectroscopic ellipsometry should be performed on modern glasses having visual properties of medieval glasses. As the church of the Royaumont abbey was destroyed, only a few pieces of glass were retrieved. Since comparisons can not be accurately done on so few medieval samples we resorted to a larger set of medieval glass samples obtained from the Maubuisson abbey which is very close to Royaumont abbey in terms of geographic location and construction time. With all these medieval glass samples, including those from the Royaumont abbey, an exhaustive collection of equivalent modern glasses from Saint Just Corporation satisfying Fresnel conditions can be determined in order to estimate the spectral absorption property. The equivalence is relative to a visual matching and is determined using a light booth, under controlled view conditions (standard illuminants). On another hand, in order to simulate the internal scattering behavior of the medieval glasses (due for example to the presence of micro-bubbles), a map of spectral transmittance is generated through a set of measurements effected at several points of interest on the medieval samples. This map is accessed by texture coordinates mechanism, and is very accurate for simulating realistic visual appearance when a large number of points are selected for the measurements. More details can be found in~\cite{magoules:proceedings-auth:46}.

A second physical aspect we considered in our Virtuelium rendering software is the properties of our scene lighting source: a representation of the sun. Even when passing over irregular fluctuations of the sun activity, its average emission spectrum should stay subject to evolution over time. On this point, two factors can be meaningful:
\begin{itemize}
\item Hours of the day: because of the Earth rotation around its axis, the sun position in the sky varies, resulting in different incident angles. With the influence of optical properties of Earth atmosphere, different incident angles lead to offsets in the emission spectrum (sunset emission spectra are more enriched in red and infrared wavelengths and weakened in blue ones).
\item Dates of the year: since the Earth revolution trajectory is an ellipse, the distance between the planet and the sun is not constant over a year, resulting in different scattering conditions by the atmosphere, leading to spectral variations, power, and incident angles.
\end{itemize}
However, in order to obtain accurate interpolation over the day and the year, we might need a rich database of measurements of the solar emission spectrum. While such a work is still in progress, we made the additional hypothesis that solar radiation can be seen from the surface of the Earth as a plane wave (instead of spherical) because distances between the sun and the Earth are large enough to be considered as infinite. We then used the Rayleigh scattering theory~\cite{bruno2002classical} to compute a plausible solar emission spectrum.

\section{Global illumination rendering}

In~\cite{Kajiya:1986:RE:15886.15902}, the author stated the following equation based on the radiative transfer equation, for modeling the interaction between light rays and objects in a three dimensional scene:
\begin{equation}
\label{eq:radiative_theo}
L_{r}(\vec{\omega_{o}}) = \int_{\Omega} F_{r}(\vec{\omega_{i}}, \vec{\omega_{o}}) L_{i}(\vec{\omega_{i}}) \vec{n}.\vec{\omega_{i}} d \omega_{i}
\end{equation}
where each $L_{i}(\vec{\omega_{i}})$ is an incident light ray emitted by a dome $\Omega$ in the direction $\vec{\omega_{i}}$. $L_{r}(\vec{\omega_{o}})$ is the remitted light in a direction $\vec{\omega_{o}}$, and $\vec{n}$ denotes the unit normal vector of the object surface. $F_{r}$ is the reflectance behavior of the object material. Transmittance property can be applied for window glasses by considering an entire sphere instead of a dome. By simplifying the equation~(\ref{eq:radiative_theo}), for $N$ directional light-sources $s=1,\ldots,N$, each one characterized by an emission spectrum $L_{s}$ and an incident direction $\vec{\omega_{s}}$, we compute the remitted light as follows (see also~\cite{magoules:proceedings-auth:64}):
\begin{equation}
\label{eq:radiative_appli}
L_{r}(\vec{\omega_{o}}) = \sum_{s=1}^N F_{r}(\vec{\omega_{s}}, \vec{\omega_{o}}) L_{s}(\vec{\omega_{s}}) \vec{n}.\vec{\omega_{s}}
\end{equation}
At last, a photon mapping algorithm~\cite{Jensen:1996:GIU:275458.275461, Jensen:2004:PGG:1103900.1103920} is used to perform a Global Illumination (GI) rendering. In GI rendering, it is not only the light sources which are responsible of the illumination of an object but also all the reflective objects around. It is an observable progress in terms of virtual reality, and many GI algorithms have been developed and improved~\cite{Wallace:1987:TSR:37402.37438, Sillion:1989:GTM:74333.74368, CGF:CGF1863, Hachisuka:2008:PPM:1409060.1409083}. Our photon mapping algorithm proceeds in two sequential steps. First, photon maps structures are filled with the position where light rays (photons) thrown up from light-sources hit the objects. The algorithm requires at least one photon map for later computing the global illumination and one for caustics. Secondly, considering the fact that the remitted light $L_{r}(\vec{\omega_{o}})$ can be a sum of separated integrals, we evaluate direct and specular contributions, then we deduce caustic and indirect diffuse contributions from photon maps. Direct and specular contributions (local illumination) are computed by the Scanline rendering algorithm~\cite{Wylie:1967:HPD:1465611.1465619}. It performs an inverse ray tracing~\cite{Arvo86backwardray} where pixels of the image are evaluated line by line, from top to bottom. This way, objects are sorted by being projected onto the image plane and considered when they contribute to the evaluation of the color of a pixel. Also, there exist progressive photon mapping algorithms~\cite{Hachisuka:2008:PPM:1409060.1409083, Hachisuka:2011:RAP:2019627.2019633} which do not limit the number of rebounds light-rays can perform. For instance, their convergence criterion can be a user setting, or based on a quality status.

\section{Fast high quality rendering}

A high quality rendering based on physical properties can last several hours since the accuracy depends on several criteria and on the number of light-rays considered. A visual appearance simulation of a whole church edifice would be practically an endless process. On another side, virtual reality or real-time interactive rendering would only be possible with some loss of accuracy (see e.g.~\cite{magoules:proceedings-auth:59}).

Our goal is to have a high quality rendering. In addition to the realistic CAD model and physically-based rendering we have defined, more than one billion light-rays are used for the simulation. A significant acceleration can be achieved by means of parallel computing. Generally, parallel image rendering consists to split the image grid in order to compute several pixels color at a time without any communication between processing units. Static load balancing distributes the whole set of pixels over the involved processors once and for all, while dynamic load balancing assigns a new subset of pixels to a processor as soon as it goes under a workload threshold. This dynamic aspect can be necessary when certain parts of the image induce more calculation time than others. Parallelization could be more efficient if one focuses on distributing the light-rays management, as it is the most costly part of the rendering process. Anyway, since light-rays could possibly spread everywhere inside the building, it would be necessary to load the whole model (including physical data) on each processor. This could be a limitation for a full utilization of multi-core architectures, due to memory availability. Other details are further discussed in~\cite{magoules:proceedings-auth:49, magoules:journal-auth:60}

Domain Decomposition Methods (DDM)~\cite{SmithBjorstadGropp1996,TW2005,Jar2007,magoules:journal-auth:21} allow to distribute the model itself without duplications. Interface tuning techniques~\cite{magoules:journal-auth:16} may be used to improve the efficiency of these methods. Continuous optimization approach first introduced in \cite{NatafEtAl1994} and then extended to other equations in \cite{CN1998,GanderHalpernNataf2003,magoules:journal-auth:23,magoules:journal-auth:24,magoules:journal-auth:18,magoules:journal-auth:14} are very efficient. An alternative which do not require the knowledge of the equations but only the matrices has been introduced in \cite{magoules:journal-auth:8,magoules:proceedings-auth:6,magoules:journal-auth:20,magoules:journal-auth:17,magoules:journal-auth:12}. The recent extension of this discrete approach to ray-tracing allows to distribute the model itself without duplications because when a light-ray propagates out of the part of the model assigned to a processor, its properties can be communicated to the adequate processor through interface conditions. We combined such DDM with a particular dynamic load balancing scheme where computation and sub-model loading and unloading are overlapped. A sub-model is unloaded when almost all the propagating light-rays are handled. But prior to the unloading, another available sub-model containing more unprocessed light-rays is loaded. This way, the processor switches from one sub-model to another without stopping light-rays processing. Further details can be found in~\cite{magoules:patent:2011, magoules:proceedings-auth:39, magoules:proceedings-auth:49, magoules:proceedings-auth:60}.

Additionally, the processing of a sub-domain itself is accelerated through a shared memory multi-threading parallelization coupled with a work-sharing load balancing scheme. While there is neither data replication nor local outputs gathering, a thread-safe access mechanism is required. Considering that the multi-threaded beam-tracing generates few contention, thus the busy-waiting time is insignificant, we resort to a simple fine-grained locking by setting a spin-lock mutex on each cell of the output array. This offers a good trade-off between simplicity and efficiency.

\section{Results and discussion}

\begin{figure}[h!]
\centering
\scalebox{0.5}{\includegraphics[width=1\textwidth]{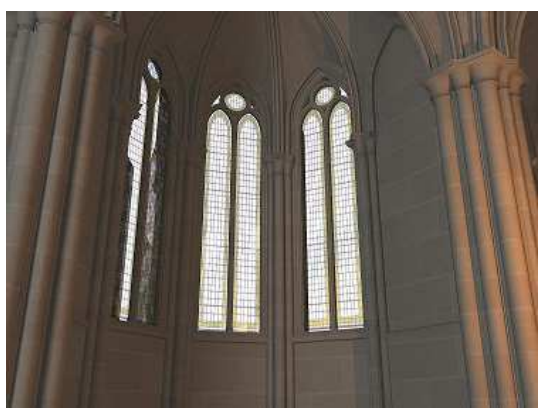}}
\caption{Close interior view of the Royaumont church - Physically-based rendering}
\label{fig:closevirtuelium}
\end{figure}

In table~\ref{tab:ddm}, we present some simulation times obtained with our DDM parallelization by rendering images like the one in fig.~\ref{fig:closevirtuelium}. On the left of this image, we can see some objects reflecting on the glass window. Unlike classical rendering, we applied spectral textures based on optical constants (complex index of refraction). It is a well known colorimetry technique by which a unique RGB data can be deduced from a given spectrum input.

On efficiency aspect, we can notice an important time saving when the proposed DDM is applied. The number of threads denotes a classical parallelization of the processing of one sub-domain (sub-model). From a sequential processing that lasted about 1 hour and 30 minutes, we fell below 3 minutes either with 64 threads on 1 sub-domain or with 32 threads (half) on 2 sub-domains (double). While there was the limit for the classical parallelization, including the domain decomposition approach allowed to keep reducing the execution time down to 1 minute.

\begin{table}[h!]
\centering
{\small
\begin{tabular}{|l|c|c|c|c|}
\hline
 & 16 & 32 & 64 & 128 \\
 & threads & threads & threads & threads \\
\hline
{1 domain}  & 5.61 & 3.57 & 2.97 & 3.88 \\
{2 sub-domains} & 5.52 & 2.93 & 1.82 & 1.59 \\
{4 sub-domains} & 5.57 & 3.06 & 1.77 & 1.66 \\
{8 sub-domains} & 5.54 & 3.08 & 1.70 & 1.07 \\
\hline
\end{tabular}
}
\caption{Execution time (in minutes) of the physically-based rendering program upon the number of threads and the number of sub-domains. Experiments have been run on a cluster of PCs equipped with a Gigabyte Ethernet network.}
\label{tab:ddm}
\end{table}

\begin{figure}[h!]
\centering
\scalebox{0.5}{\includegraphics[width=1\textwidth]{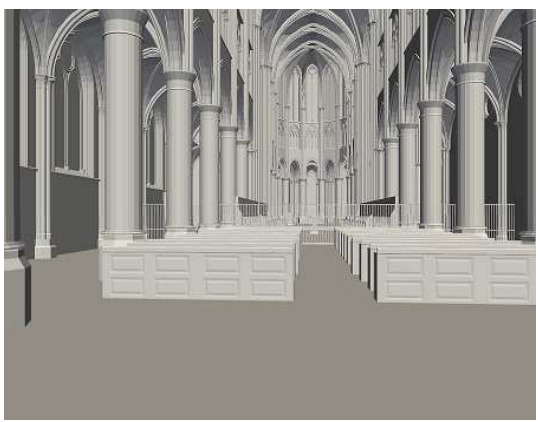}}
\caption{CAD model of the Royaumont church (global interior view)}
\label{fig:globmodel}
\end{figure}

Figure~\ref{fig:globmodel} illustrates the three dimensional CAD model from a camera object capturing the interior of the virtual church.

\begin{figure}[h!]
\centering
\scalebox{0.5}{\includegraphics[width=1\textwidth]{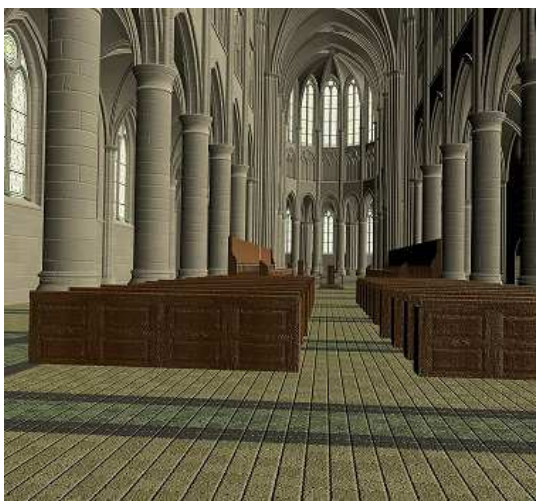}}
\caption{Basic texture rendering of the Royaumont church (global interior view)}
\label{fig:globtexture}
\end{figure}

Figure~\ref{fig:globtexture} shows the interior of the virtual church, rendered with the Blender software. Beyond the visible geometrical model, simple texture images were mapped on the different objects.

\begin{figure}[h!]
\centering
\scalebox{0.5}{\includegraphics[width=1\textwidth]{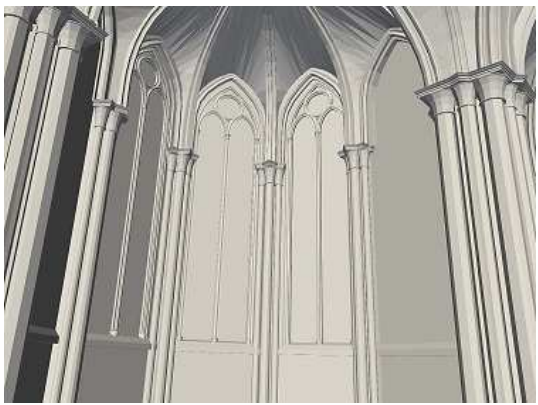}}
\caption{Close interior view of the Royaumont church - CAD model}
\label{fig:closemodel}
\end{figure}

\begin{figure}[h!]
\centering
\scalebox{0.5}{\includegraphics[width=1\textwidth]{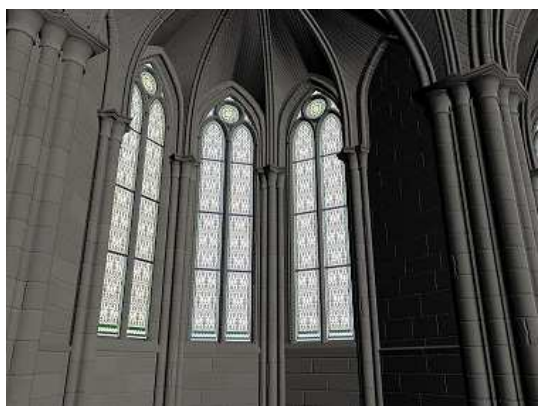}}
\caption{Close interior view of the Royaumont church - Basic texture rendering}
\label{fig:closetexture}
\end{figure}

Figures~\ref{fig:closemodel} to~\ref{fig:closevirtuelium} show close views comparing respectively the CAD model, a basic texture rendering obtained with the Blender software and a physically-based rendering obtained with Virtuelium including the proposed parallel domain decomposition method. We can see lighting effects inside the church and reflections on a window glass. Textures are made up according to spectral data which are mapped to RGB images by means of colorimetry. The properties of the window glasses match clearness criteria of the Cistercian art.

\section{Conclusion}

Catholic religious edifices are of great interest in the cultural heritage of France, as this religion mainly influenced the socio-political organization of the realm for many centuries. As a part of a rehabilitation project led by the Fondation Royaumont, a virtual reconstitution of the destroyed church of Royaumont is being realized. A first step consisted in making up a three dimensional computer aided design model representing the church in a very detailed way. Historians and archaeologists actively contributed with us to the designing of the non-existent edifice. Then, for achieving realistic visual aspects of the interior of the church, taking into account natural lighting and window glasses properties, a physically-based rendering engine was developed. Input physical data were measured on a few remains of the edifice and on samples coming from other similar churches.

As a high quality rendering lasts several hours, our aim was to accelerate the rendering process. We coupled DDM techniques with a classical parallelization of the algorithm, while observing an adapted dynamic load balancing. According to computational time observed with several renderings of the church interior, it appeared that the DDM approach significantly improved the efficiency of the parallelization.

\section*{Acknowledgements}

To conduct this project a scientific committee composed of Art Historians, Architects, Archaeologists, Mathematicians, and Computer Scientists has been formed.
The authors acknowledge (in alphabetical order), Jean-Louis Bernard (INRAP), Damien Bern\'e (Mus\'ee de Cluny à Paris), R\'emi Cerise (D2C), J\'er\^ome Johnson (Fondation Royaumont), Sophie Lagrabrielle (Mus\'ee National du Moyen Age), Jean-Michel Leniaud (Ecole Nationale des Chartes), Nathalie Le Gonidec (Fondation Royaumont), Pierre-Yves Le Pogam (Mus\'ee du Louvre), Claire Ma\^itre (CNRS), Francis Mar\'echal (Fondation Royaumont), Philippe Plagnieux (Universit\'e de Franche-Comt\'e), Marc Vir\'e (INRAP), and Monique Wabont (Val-d'Oise), for the useful discussions making this project possible.

\bibliography{ref}

\begin{thebibliography}{10}

\bibitem{Arvo86backwardray}
J.~Arvo.
\newblock Backward ray tracing.
\newblock In {\em In ACM SIGGRAPH ’86 Course Notes - Developments in Ray
  Tracing}, pages 259--263, 1986.

\bibitem{add1}
M.~Berger, A.~Tagliasacchi, L.~Seversky, P.~Alliez, J.~Levine, A.~Sharf, and
  C.~Silva.
\newblock State of the art in surface reconstruction from point clouds.
\newblock In {\em Proceedings of Eurographics 2014 - State of the Art Reports,
  Vol. 1 of EUROGRAPHICS star report, Strasbourg, France}, pages 161--185,
  2014.

\bibitem{bruno2002classical}
G.~Bruno.
\newblock Classical theory of rayleigh and raman scattering.
\newblock {\em The Raman Effect: A Unified Treatment of the Theory of Raman
  Scattering by Molecules}, pages 31--48, 2002.

\bibitem{callet1998couleur}
P.~Callet.
\newblock {\em Couleur-lumi{\`e}re, couleur-mati{\`e}re: interaction
  lumi{\`e}re-mati{\`e}re et synth{\`e}se d'images}.
\newblock Sciences en actes. Diderot {\'e}diteur, Arts et sciences, 1998.

\bibitem{magoules:proceedings-auth:46}
R.~Cerise, F.~da~Graca, F.~Magoul\`es, and P.~Callet.
\newblock Natural lighting and medieval glass: Scientific data acquisition,
  methodology and physically based rendering.
\newblock In {\em Proceedings of the International Conference on Cultural
  Heritage (EuroMed 2012), Lemesos, Cyprus, October 29--November 3, 2012},
  volume 7616. {Lecture Notes in Computer Science (LNCS), Springer-Verlag},
  2012.

\bibitem{CN1998}
P.~Chevalier and F.~Nataf.
\newblock Symmetrized method with optimized second-order conditions for the
  {H}elmholtz equation.
\newblock In {\em Domain decomposition methods, 10 (Boulder, CO, 1997)}, pages
  400--407. Amer. Math. Soc., Providence, RI, 1998.

\bibitem{GanderHalpernNataf2003}
M.~Gander, L.~Halpern, and F.~Nataf.
\newblock Optimal schwarz waveform relaxation for the one dimensional wave
  equation.
\newblock {\em SIAM Journal on Numerical Analysis}, 41(5):1643--1681, 2003.

\bibitem{magoules:proceedings-auth:59}
G.~Gbikpi-Benissan, P.~Callet, and F.~Magoul\`es.
\newblock Spectral domain decomposition method for physically-based rendering
  of photochromic/electrochromic glass windows.
\newblock In {\em Proceedings of the 13th International Symposium on
  Distributed Computing and Applications to Business, Engineering and Science
  (DCABES), Xianning, China, November 24-27, 2014}. {IEEE Computer Society},
  2014.

\bibitem{magoules:proceedings-auth:64}
G.~Gbikpi-Benissan, R.~Cerise, P.~Callet, and F.~Magoul\`es.
\newblock Spectral domain decomposition method for natural lighting and
  medieval glass rendering.
\newblock In {\em Proceedings of the 16th International Conference on High
  Performance Computing and Communications (HPCC 2014), Paris, France,
  Aug.20-22, 2014.} {IEEE Computer Society}, 2014.

\bibitem{magoules:proceedings-auth:60}
G.~Gbikpi-Benissan and F.~Magoul\`es.
\newblock Beam-tracing domain decomposition method for urban acoustic
  pollution.
\newblock In {\em Proceedings of the 13th International Symposium on
  Distributed Computing and Applications to Business, Engineering and Science
  (DCABES), Xianning, China, November 24-27, 2014}. {IEEE Computer Society},
  2014.

\bibitem{Hachisuka:2011:RAP:2019627.2019633}
T.~Hachisuka and H.~W. Jensen.
\newblock Robust adaptive photon tracing using photon path visibility.
\newblock {\em ACM Trans. Graph.}, 30(5):114:1--114:11, Oct. 2011.

\bibitem{Hachisuka:2008:PPM:1409060.1409083}
T.~Hachisuka, S.~Ogaki, and H.~W. Jensen.
\newblock Progressive photon mapping.
\newblock {\em ACM Trans. Graph.}, 27(5):130:1--130:8, Dec. 2008.

\bibitem{Jensen:1996:GIU:275458.275461}
H.~W. Jensen.
\newblock Global illumination using photon maps.
\newblock In {\em Proceedings of the Eurographics Workshop on Rendering
  Techniques '96}, pages 21--30, London, UK, UK, 1996. Springer-Verlag.

\bibitem{Jensen:2004:PGG:1103900.1103920}
H.~W. Jensen.
\newblock A practical guide to global illumination using ray tracing and photon
  mapping.
\newblock In {\em ACM SIGGRAPH 2004 Course Notes}, SIGGRAPH '04, New York, NY,
  USA, 2004. ACM.

\bibitem{Kajiya:1986:RE:15886.15902}
J.~T. Kajiya.
\newblock The rendering equation.
\newblock {\em SIGGRAPH Comput. Graph.}, 20(4):143--150, Aug. 1986.

\bibitem{Jar2007}
J.~Kruis.
\newblock {\em Domain Decomposition Methods for Distributed Computing}.
\newblock Saxe-Coburg Publications, 2007.

\bibitem{KBKHL2004}
S.-W. Kwon, F.~Bosche, C.~Kim, C.~T. Haas, and K.~A. Liapi.
\newblock Fitting range data to primitives for rapid local 3d modeling using
  sparse range point clouds.
\newblock Technical report, Department of Civil Engineering, The University of
  Texas at Austin, Austin, TX 78712, USA, 2004.

\bibitem{magoules:journal-auth:14}
Y.~Maday and F.~Magoul\`es.
\newblock Non-overlapping additive {S}chwarz methods tuned to highly
  heterogeneous media.
\newblock {\em Comptes Rendus {\`a} l'Acad{\'e}mie des Sciences},
  341(11):701--705, 2005.

\bibitem{magoules:journal-auth:16}
Y.~Maday and F.~Magoul\`es.
\newblock Absorbing interface conditions for domain decomposition methods: a
  general presentation.
\newblock {\em Computer Methods in Applied Mechanics and Engineering},
  195(29--32):3880--3900, 2006.

\bibitem{magoules:journal-auth:18}
Y.~Maday and F.~Magoul\`es.
\newblock Improved ad hoc interface conditions for {S}chwarz solution procedure
  tuned to highly heterogeneous media.
\newblock {\em Applied Mathematical Modelling}, 30(8):731--743, 2006.

\bibitem{magoules:journal-auth:24}
Y.~Maday and F.~Magoul\`es.
\newblock Optimal convergence properties of the {FETI} domain decomposition
  method.
\newblock {\em International Journal for Numerical Methods in Fluids},
  55(1):1--14, 2007.

\bibitem{magoules:journal-auth:23}
Y.~Maday and F.~Magoul\`es.
\newblock Optimized {S}chwarz methods without overlap for highly heterogeneous
  media.
\newblock {\em Computer Methods in Applied Mechanics and Engineering},
  196(8):1541--1553, 2007.

\bibitem{magoules:patent:2011}
F.~Magoul\`es.
\newblock D\'ecomposition de domaines pour le lancer de rayons.
\newblock {F}rance {P}atent no. 1157329. 12 August 2011.

\bibitem{magoules:proceedings-auth:49}
F.~Magoul\`es, R.~Cerise, and P.~Callet.
\newblock A beam-tracing domain decomposition method for sound holography in
  church acoustics.
\newblock In {\em Proceedings of the 12th International Symposium on
  Distributed Computing and Applications to Business, Engineering and Science
  (DCABES), Kingston, London, UK, September 2nd-4th, 2013}. {IEEE Computer
  Society}, 2013.

\bibitem{magoules:journal-auth:60}
F.~Magoul\`es, G.~Gbikpi-Benissan, and P.~Callet.
\newblock Ray-tracing domain decomposition methods for real-time simulation on
  multi-core and multi-processor systems.
\newblock {\em Concurrency and Computation: Practice and Experience}, pages~--,
  (2016, in press).

\bibitem{magoules:journal-auth:21}
F.~Magoul\`es and F.-X. Roux.
\newblock Lagrangian formulation of domain decomposition methods: a unified
  theory.
\newblock {\em Applied Mathematical Modelling}, 30(7):593--615, 2006.

\bibitem{magoules:journal-auth:8}
F.~Magoul\`es, F.-X. Roux, and S.~Salmon.
\newblock Optimal discrete transmission conditions for a non-overlapping domain
  decomposition method for the {H}elmholtz equation.
\newblock {\em SIAM Journal on Scientific Computing}, 25(5):1497--1515, 2004.

\bibitem{magoules:journal-auth:12}
F.~Magoul\`es, F.-X. Roux, and L.~Series.
\newblock Algebraic way to derive absorbing boundary conditions for the
  {H}elmholtz equation.
\newblock {\em Journal of Computational Acoustics}, 13(3):433--454, 2005.

\bibitem{magoules:journal-auth:17}
F.~Magoul\`es, F.-X. Roux, and L.~Series.
\newblock Algebraic approximation of {D}irichlet-to-{N}eumann maps for the
  equations of linear elasticity.
\newblock {\em Computer Methods in Applied Mechanics and Engineering},
  195(29--32):3742--3759, 2006.

\bibitem{magoules:journal-auth:20}
F.~Magoul\`es, F.-X. Roux, and L.~Series.
\newblock Algebraic {D}irichlet-to-{N}eumann mapping for linear elasticity
  problems with extreme contrasts in the coefficients.
\newblock {\em Applied Mathematical Modelling}, 30(8):702--713, 2006.

\bibitem{Mil1791}
A.~Millin.
\newblock {\em Antiquit\'es nationales, ou Recueil de monuments pour servir
  l'Histoire}, volume~2.
\newblock 1791.

\bibitem{NatafEtAl1994}
F.~Nataf, F.~Rogier, and E.~de~Sturler.
\newblock Optimal interface conditions for domain decomposition methods.
\newblock Technical report, CMAP, Ecole Polytechnique, UMR CNRS 7641, 91128
  Palaiseau Cedex, France, 1994.
\newblock I.R. no 301.

\bibitem{CGF:CGF1863}
A.~Pajot, L.~Barthe, M.~Paulin, and P.~Poulin.
\newblock Combinatorial bidirectional path-tracing for efficient hybrid cpu/gpu
  rendering.
\newblock {\em Computer Graphics Forum}, 30(2):315--324, 2011.

\bibitem{palik1985handbook}
E.~Palik.
\newblock {\em Handbook of Optical Constants of Solids}.
\newblock Number vol.~1. Elsevier Science, 1985.

\bibitem{Pharr:2010:PBR:1854996}
M.~Pharr and G.~Humphreys.
\newblock {\em Physically Based Rendering, Second Edition: From Theory To
  Implementation}.
\newblock Morgan Kaufmann Publishers Inc., San Francisco, CA, USA, 2nd edition,
  2010.

\bibitem{magoules:proceedings-auth:6}
F.-X. Roux, F.~Magoul\`es, L.~Series, and Y.~Boubendir.
\newblock Approximation of optimal interface boundary conditions for
  two-{L}agrange multiplier {FETI} method.
\newblock In R.~Kornhuber, R.~Hoppe, J.~P\'eriaux, O.~Pironneau, O.~Widlund,
  and J.~Xu, editors, {\em Proceedings of the 15th International Conference on
  Domain Decomposition Methods, Berlin, Germany, July 21-15, 2003}, Lecture
  Notes in Computational Science and Engineering. Springer-Verlag, Haidelberg,
  2005.

\bibitem{Shirley:2012:BPR:2407783.2407785}
P.~Shirley, R.~K. Morley, P.-P. Sloan, and C.~Wyman.
\newblock Basics of physically-based rendering.
\newblock In {\em SIGGRAPH Asia 2012 Courses}, SA '12, pages 2:1--2:11, New
  York, NY, USA, 2012. ACM.

\bibitem{Sillion:1989:GTM:74333.74368}
F.~Sillion and C.~Puech.
\newblock A general two-pass method integrating specular and diffuse
  reflection.
\newblock In {\em Proceedings of the 16th Annual Conference on Computer
  Graphics and Interactive Techniques}, SIGGRAPH '89, pages 335--344, New York,
  NY, USA, 1989. ACM.

\bibitem{SmithBjorstadGropp1996}
B.~Smith, P.~Bjorstad, and W.~Gropp.
\newblock {\em Domain Decomposition: Parallel Multilevel Methods for Elliptic
  Partial Differential Equations}.
\newblock Cambridge University Press, 1996.

\bibitem{add2}
P.~Tang, D.~Huber, B.~Akinci, R.~Lipman, and A.~Lytle.
\newblock Automatic reconstruction of as-built building information models from
  laser-scanned point clouds: A review of related techniques.
\newblock {\em Automation in Construction}, 19(7):829--843, 2010.

\bibitem{TW2005}
A.~Toselli and O.~Widlund.
\newblock {\em Domain Decomposition methods: Algorithms and Theory}.
\newblock Springer, 2005.

\bibitem{magoules:proceedings-auth:39}
C.~Venet and F.~Magoul\`es.
\newblock Parallel domain decomposition methods for ray-tracing on multi-cores
  and multi-processors.
\newblock In {\em Proceedings of the 10th International Symposium on
  Distributed Computing and Applications to Business, Engineering and Science
  (DCABES), Wuxi, China, October 14--17, 2011}. {IEEE Computer Society}, 2011.

\bibitem{Wallace:1987:TSR:37402.37438}
J.~R. Wallace, M.~F. Cohen, and D.~P. Greenberg.
\newblock A two-pass solution to the rendering equation: A synthesis of ray
  tracing and radiosity methods.
\newblock {\em SIGGRAPH Comput. Graph.}, 21(4):311--320, Aug. 1987.

\bibitem{Wylie:1967:HPD:1465611.1465619}
C.~Wylie, G.~Romney, D.~Evans, and A.~Erdahl.
\newblock Half-tone perspective drawings by computer.
\newblock In {\em Proceedings of the November 14-16, 1967, Fall Joint Computer
  Conference}, AFIPS '67 (Fall), pages 49--58, New York, NY, USA, 1967. ACM.

\end{thebibliography}
\bibliographystyle{abbrv}

\end{document}